\documentclass[prb,twocolumn,showpacs,preprintnumbers,amsmath,amssymb,floatfix]{revtex4}
\usepackage{hyperref}
\usepackage{epsfig}
\usepackage{graphicx}
\usepackage{dcolumn}
\usepackage{bm}
\usepackage{amsfonts}
\usepackage{amsmath}

\begin{document}

\title{Optimized Cooper pair pumps}
\date{\today}

\author{Shabnam Safaei}
\email{safaei@sns.it}
\affiliation{NEST-CNR-INFM and Scuola Normale Superiore, Piazza dei Cavalieri 7, I-56126 Pisa, Italy}
\author{Simone Montangero}
\affiliation{NEST-CNR-INFM and Scuola Normale Superiore, Piazza dei Cavalieri 7, I-56126 Pisa, Italy}
\author{Fabio Taddei}
\affiliation{NEST-CNR-INFM and Scuola Normale Superiore, Piazza dei Cavalieri 7, I-56126 Pisa, Italy}
\author{Rosario Fazio}
\affiliation{International School for Advanced Studies (SISSA), Via Beirut 2-4, I-34014 Trieste, Italy}
\affiliation{NEST-CNR-INFM and Scuola Normale Superiore, Piazza dei Cavalieri 7, I-56126 Pisa, Italy}

\begin{abstract}
In adiabatic Cooper pair pumps, operated by means of gate voltage modulation only, the 
quantization of the pumped charge during a cycle is limited due to the quantum coherence of 
the macroscopic superconducting wave function. In this work we show that it is possible to 
obtain very accurate pumps in the non-adiabatic regime by a suitable choice of the shape of 
the gate voltage pulses. We determine the shape of these pulses by applying quantum optimal 
control theory to this problem. In the optimal case the error, with respect to the quantized 
value,  can be as small as of the order of $10^{-6}e$: the error is
reduced by up to five orders of magnitude with respect to the
adiabatic pumping. 
In order to test the experimental 
feasibility of this approach we consider the effect of charge noise and the deformations
of the optimal pulse shapes on the accuracy of the pump. Charge noise
is assumed to be induced by random background charges in the substrate, responsible for the observed 
$1/f$ noise. Inaccuracies in the pulse shaping are described by assuming a finite bandwidth 
for the pulse generator. In realistic cases the error increases at
most of one order of magnitude as compared to the optimal case. Our results are promising for 
the realization of accurate and fast superconducting pumps.    
\end{abstract}
\maketitle

\section{Introduction}
A dc current can be generated in  a mesoscopic circuit connected to two electrodes even in the 
absence of a bias voltage if a  set  of external  parameters, e.g. gate voltages, are 
changed periodically in time~\cite{thouless83}. This mechanism is known as charge pumping. If 
the external parameters are changed slowly as compared to the typical  time  scales of the system the 
pumping is adiabatic. In this case the charge pumped after one cycle does not depend on the detailed 
timing of the cycle, but only on its geometrical properties. Charge pumping can be realized in 
a variety of different situations. In systems consisting of mesoscopic phase-coherent conductors 
parametric pumping 
can be achieved through a periodic modulation of the phases of the scattering matrix.  This regime has been 
studied extensively both for metallic conductors, see
e.g. Refs.~\onlinecite{brouwer98,zhou99,makhlin01,moskalets02a,moskalets02b,
entin02} and  hybrid systems~\cite{wang01,blaauboer02,taddei04} containing 
superconducting terminals.  In the opposite regime of systems consisting of quantum dots connected 
through tunnel junctions, charge pumping is achieved by the periodic modulation of the Coulomb 
blockade~\cite{SCT92}. In this case  phase coherence is irrelevant 
and the number of electrons transfered per cycle is approximately quantized: 
the generated current $I$ is related to the frequency of cycle $f$ via $I \sim (ne)f$, where 
$n$ is the number of electrons transfered in each cycle and $e$ is the electron charge. Experimental 
evidence for parametric charge pumping in normal metallic systems has been demonstrated for the first time 
in Refs.~\onlinecite{kouwenhoven91,pothier92}.

The situation is radically different if superconducting islands are considered. Here at low 
temperatures pumping is due to the transport of Cooper  pairs~\cite{geerligs91,niskanen05}. A Cooper 
pair pump can be realized by an array of Josephson junctions connected to two superconducting 
reservoirs, kept at a fixed phase bias $\varphi$~\cite{footnote}. In this case, even in those situations 
where pumping is associated to a periodic modulation of  the Coulomb  blockade, superconducting phase 
coherence is fundamental~\cite{pekola99,niskanen03}. Moreover, in addition to the dependence of the pumped 
charge on the characteristics of the cycle, in superconducting pumps there is a dependence on the 
superconducting phase difference between the reservoirs.  The geometric nature of the pumped charge 
has been analyzed both in the Abelian~\cite{aunola03,governale05,mottonen06,leone07} and 
non-Abelian~\cite{brosco07} cases thus opening the possibility to experimentally detect geometric 
phases in superconducting nanocircuits~\cite{falci00, leek07}. Indeed an experimental detection of the Berry phase 
by means of Cooper pair pumping has recently been reported~\cite{mottonen07}.  

\begin{figure}
\begin{center}
\vspace*{1cm}
\includegraphics[width=0.6\linewidth]{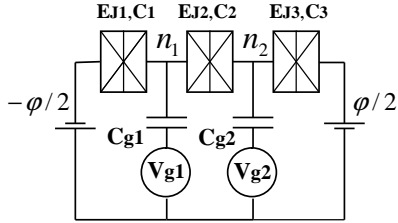}
\caption{Schematic drawing of a Cooper pair pump consisting of three Josephson junctions and two 
         gated Cooper pair boxes. $C_{i}$ and $E_{Ji}$ are the capacitances and Josephson energies of the 
         junctions. $C_{gi}$ and $V_{gi}$ are gate capacitances and voltages, $\varphi$ is the overall phase 
         difference. By a proper modulation of the gate voltages a given charge is moved from, say, the 
         left to the right electrode. In general the pumped charge will depend on the modulation shape and 
         on the phase difference $\varphi$. In  the protocol considered here the dependence on $\varphi$ is 
         absent.
\label{fig:pump}}
\end{center}
\end{figure}
In addition to the importance of addressing fundamental questions related to the quantum mechanical 
behavior of macroscopic systems, accurate charge pumping can be used for metrological purposes. In the case 
of single-electron pumps the transfered charge, at frequency $f$ of a few MHz, has reached such an 
accuracy (uncertainty of $10^{-8}$) to make a new metrological standard of capacitance 
possible~\cite{keller96, keller98, keller99}. By pumping a certain number of electrons onto a capacitor, 
and measuring the resulting voltage, one can measure capacities of the order of 1pF.  On the contrary, 
the frequencies, and consequently the current intensities, in single-electron pumps has not been 
sufficiently high for setting a standard of current. A superconducting Cooper pair pump, in principle, 
would allow for higher frequencies although several  effects such as Landau-Zener tunneling, supercurrent 
leakage through the pump, and coherent corrections (which are of crucial importance to reveal geometric 
phases)  would lead to all sorts of inaccuracies in Cooper pair pumping. Some works have been done to 
achieve more accurate pumping in adiabatic regime by optimizing the design of the pump~\cite{niskanen03, 
cholascinski07,leone07}. In this paper we follow a different approach.
We will design very accurate Cooper pair pumps by optimizing the pulse shapes of the gate voltages 
during the cycle. Most importantly we are not bound to work in the adiabatic limit therefore increasing,
at the same time,  both the accuracy and the frequency of operation. The theoretical framework which 
will apply to achieve this goal is that of  optimal control theory~\cite{peirce88,borzi002,sklarz002,calarco04}. 
This approach was successfully applied to superconducting qubits to improve one- and two-qubit 
gates~\cite{sporl07,montangero07}  and it was shown that it can help in reducing the error in the gate 
operation by several  order of magnitudes.  As we will describe in the rest of the paper, optimal 
control is also helpful in the case of Cooper pair pumping where, for the case we consider, the error 
is reduced up to five orders of magnitude.

The paper is organized as follows. In the next Section we will describe the model for the Cooper 
pair pump used in the rest of the paper. We will then give (Sec.~\ref{oqc}) a brief  introduction 
to the optimal control algorithm employed in this work. In Sec.~\ref{res} we present the results 
obtained for our non adiabatic optimal pump.
After having discussed the achieved accuracy (Sec.~\ref{optimal}) we analyze in details the effect of 
possible imperfections in the pulse shapes (Sec.~\ref{shapes}), the effect of noise (Sec.~\ref{noise}),
as well as the effect of unavoidable variations in charging and Josephson energy from junction to junction (Sec.~\ref{parameters}),
 in order to test the robustness of the pump. The last Section is devoted to the conclusions and to a comparison 
with other proposals to realize accurate pumps.\\

\section{The Cooper pair pump}
\label{model}

The three-junction Cooper pair pump considered in the paper is shown in Fig.~\ref{fig:pump}. 
It consists of an array of three Josephson junctions with two gated islands; $E_{Ji}$ and $C_i$ are, 
respectively, Josephson energy and capacity of $i$th junction, while $C_{gi}$ and $V_{gi}$ are, respectively, 
gate capacitance and gate voltage applied on the $i$th island. 
The number of extra Cooper pairs on the $i$th island is denoted with $n_i$ and $\varphi$ is the overall 
phase difference between the two superconducting electrodes. In the case of a uniform array, which we consider
here, the Hamiltonian, in the basis of charge eigenstates $|\stackrel{\rightarrow}{n}\rangle$, is given 
by~\cite{pekola99}
\begin{widetext}
\begin{eqnarray}
\label{eq:H3}
H =  \sum_{\stackrel{\rightarrow}{n}}\left\{\frac{2}{3} E_{C} \left[(n_{1}-q_{1})^2 + (n_{2}-q_{2})^2 + 
        (n_{1}-q_{1})(n_{2}-q_{2})\right]  | \stackrel{\rightarrow}{n} \rangle \langle \stackrel{\rightarrow}{n} |
       - \frac{E_{J}}{2}  \sum^{3}_{k=1} (e^{i\frac{\varphi}{3}} |\stackrel{\rightarrow}{n}+
       \stackrel{\rightarrow}{\delta_{k}}  > 
       < \stackrel{\rightarrow}{n}| + h.c.)\right\}
\end{eqnarray}
\end{widetext}
In this equation $\stackrel{\rightarrow}{n}=(n_{1}, n_{2})$ and $\stackrel{\rightarrow}{q}=(q_{1}, q_{2})$ 
specify, respectively, the number of Cooper pairs on each island and the normalized gate charges 
$(q_{k}=-C_{g,k}V_{g,k}/(2e))$. Tunneling of one Cooper pair through the $k$th junction changes the number of
 pairs on $k$th and $(k-1)$th island, so that the only non-zero components of $\stackrel{\rightarrow}{\delta_{k}}$ 
are $(\stackrel{\rightarrow}{\delta_{k}})_{k}=1$ and $(\stackrel{\rightarrow}{\delta_{k}})_{k-1}=-1$. Moreover 
the forward (backward) tunneling of one pair through each junction is related to a $\frac{\varphi}{3}$ 
($-\frac{\varphi}{3}$) phase difference. The pump operates in the charging limit, i.e. $E_J \ll E_C$.
\begin{figure}
\begin{center}
\includegraphics[width=0.8\linewidth]{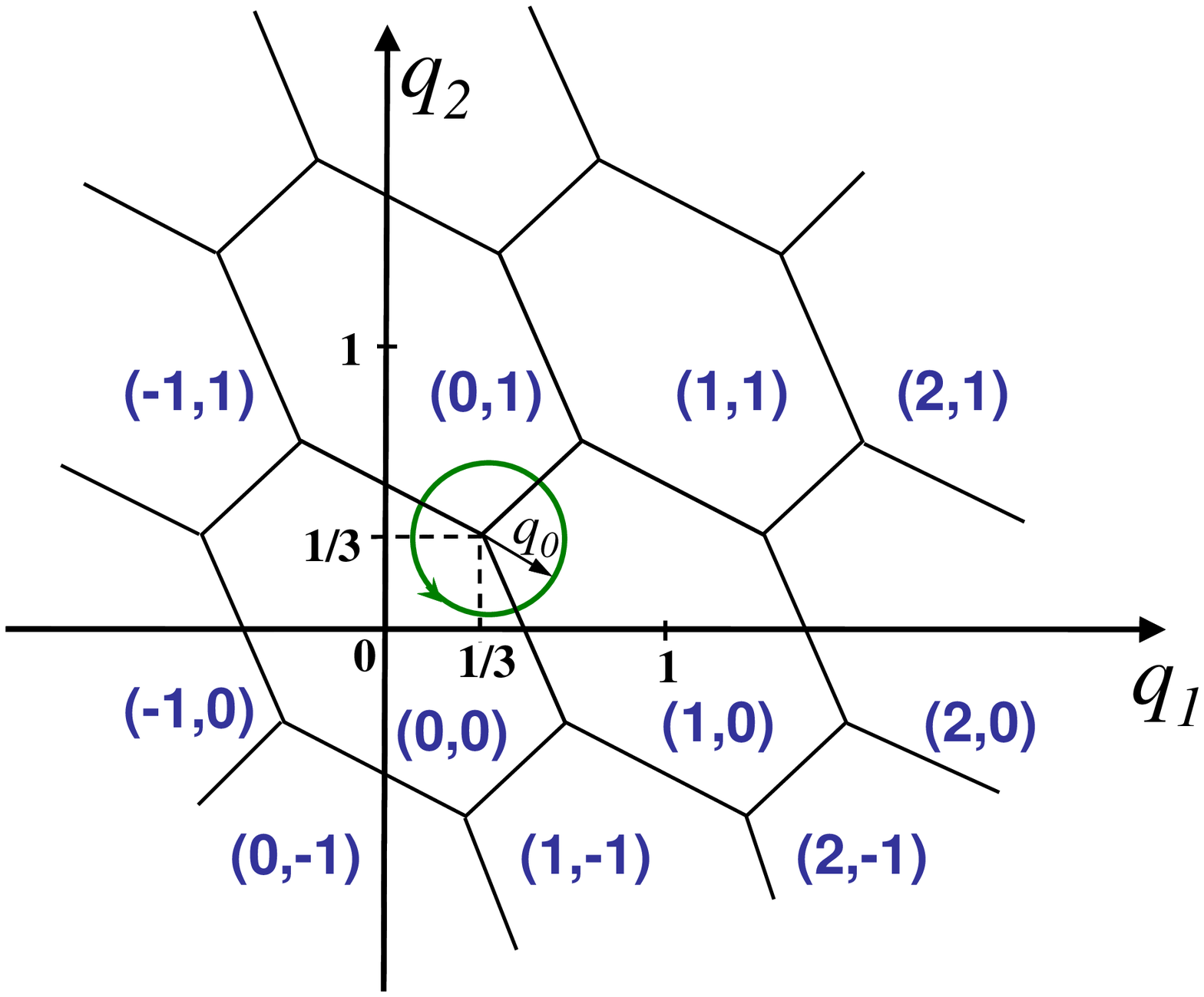}
\caption{(Color on line) The stability diagram of a three-junction Cooper pair pump in the $(q_1,q_2)$ space. 
         Each hexagon corresponds to a given charge (eigen)state  (for $E_J=0$) of the system 
         $(n_1,n_2)$. In the adiabatic regime charge is pumped when the gate charges are varied along, 
         for example, the circular path with radius $q_0$ centered at the 
         degeneracy point ($q_1=q_2=1/3$). 
\label{fig:stability}}
\end{center}
\end{figure}
As discussed in Ref.~\onlinecite{pekola99} the Cooper pair pump is operated by changing periodically in time the two 
gate charges $q_1$ and $q_2$.

It is useful to first analyze the case in which the pump is driven adiabatically. In this case
the time dependence of the gate charge is not important, what matters is the path which is followed in the 
parameter space. As was discussed extensively in the past (see
e.g. Ref.~\onlinecite{geerligs91,pekola99}) it is important 
that the cycle encloses the triple point degeneracy ($q_1=q_2=1/3$) in the stability diagram as described 
in Fig.\ref{fig:stability}. The various regions in Fig.\ref{fig:stability} indicate the corresponding ground 
state of the charging part of the Hamiltonian.  After a cycle, for example  the one indicated in the figure, 
the state of the pump goes back to its initial situation but (approximately) one Cooper pair has been 
transferred through the system. As it has been discussed in~\cite{pekola99}, this pump does not lead to 
a sufficient quantization accuracy: the error scales as $E_J/E_C$. A smaller error can be achieved either by 
increasing the number of junctions or by pumping by means of gate voltage and flux modulation as in the 
Cooper pair sluice~\cite{niskanen03}.

In the present work we take a different approach to optimize the pump. Our proposed device operates in the 
non-adiabatic limit, in which the time dependence of the pulses is
important. In order to have an 
accurate pump we will then optimize the pulse shapes by means of quantum optimal control (see next Section).
These ingredients allow to construct  a {\em fast} and  {\em accurate} Cooper pair pump.  

As we will show in the next section, in order to use the quantum optimal control, one needs to have a desired 
final state differing from the initial state and at least one parameter as a control in the Hamiltonian of the 
system. Since in pumping the initial and final states of the system are the same apart from a phase we introduce 
a new quantum number, the counter $| m \rangle$. This counter is a passive detector which 
clicks each time a Cooper pair goes through the junction where the counter is set. In the following we assume 
that the counter is placed in the  last junction of the array, therefore if one pair passes this junction from 
left (right) to right (left) the counter will change by 1 ($-1$).
 We emphasize that the counter considered is merely a fictitious element which we use to compute the pumped charge.
  No back action can be expected since we are not describing a physical detector.
  If one denotes the state of the array by $|\Phi(t)\rangle$ and the state of the counter by $|m\rangle$, 
  where $m=0, \pm1, \pm2 ...$, the state of the system will be $|\Phi(t)\rangle\otimes|m\rangle$.
 In the Hilbert space of charge states plus 
the counter index the Hamiltonian of a three-junction uniform pump has the following form:
\begin{widetext}
\begin{eqnarray}
\label{eq:H3m}
        \cal H &=& \frac{2}{3} E_{C} \sum_{m,\stackrel{\rightarrow}{n}} 
        \left[(n_{1}-q_{1})^2 + (n_{2}-q_{2})^2 + (n_{1}-q_{1})(n_{2}-q_{2})\right] 
        | \stackrel{\rightarrow}{n},m \rangle \langle \stackrel{\rightarrow}{n},m | \nonumber\\ 
        &-& \frac{1}{2} E_{J} \sum_{m,\stackrel{\rightarrow}{n}} \left[e^{i\frac{\varphi}{3}} 
        \left(\left|n_{1}+1, n_{2}, m\right\rangle
        + \left|n_{1}-1, n_{2}+1, m\right\rangle + \left|n_{1}, n_{2}-1, m+1\right\rangle \right)
        \left\langle n_{1}, n_{2}, m\right| + h.c. \right]
\end{eqnarray}
\end{widetext}
As it is clear in the Hamiltonian~(\ref{eq:H3m}), one can change the state of the counter without changing 
the energy so that all states $|\Phi(t)\rangle\otimes|m\rangle$ with
different values of $m$ are degenerate. 

\section{Optimal quantum control}
\label{oqc}

Quantum optimal control algorithms~\cite{peirce88, borzi002, sklarz002,calarco04} are designed to lead a quantum 
system with state $|\psi(t)\rangle$ from an initial state $|\psi(0)\rangle=|\psi_{ini}\rangle$ to a target final 
state $|\psi_{fin}\rangle$ at time $T$ with a good fidelity $\cal F$, where ${\cal F}\equiv|\langle \psi_{fin} | 
\psi(T)\rangle|^{2}$.  The algorithm employed in this work is called \textit{immediate feedback control} and is 
guaranteed to give a fidelity improvement at each iteration~\cite{ref:Sola98}. The procedure is as follows: Assume 
that $\{u_{j}(t)\}$ is the set of control parameters which the Hamiltonian depends on. First the state 
of the system $|\psi(t)\rangle$ is evolved in time with the initial condition $|\psi(0)\rangle=|\psi_{ini}\rangle$ 
and an initial guess $\{u^{(0)}_{j}(t)\}$ for control parameters, giving rise to $|\psi(T)\rangle$ after time $T$. 
At this point an iterative algorithm starts, aiming at improving the fidelity by adding a correction to control 
parameters in each step. In the $n$th step of this iterative algorithm 
\begin{itemize}
\item The auxiliary state $|\chi(T)\rangle\equiv |\psi_{fin}\rangle\langle\psi_{fin}|\psi(T)\rangle$ is evolved 
backwards in time reaching $|\chi(0)\rangle$. $|\chi(T)$ can be interpreted as the part of the desired final 
state $|\psi_{fin}\rangle$ which has been reached. 
\item The states $|\chi(0)\rangle$ and $|\psi(0)\rangle$ are evolved forward in time with control parameters 
$\{u^{(n)}_{j}(t)\}$ and $\{u^{(n+1)}_{j}(t)\}$, respectively. Here,
\begin{eqnarray}
\label{eq:update}
              u^{(n+1)}_{j}(t) = u^{(n)}_{j}(t) + \frac{2}{\lambda(t)} \Im\left[\langle\chi(t)|\frac{\partial H}
                {\partial u_{j}(t)}|\psi(t)\rangle\right]
\end{eqnarray}
are updated control parameters and $\lambda(t)$ is a weight function used to fix initial and final conditions 
on the control parameters.
\end{itemize}
These two steps are repeated until the desired value of fidelity is obtained.

In this work we will use the described quantum optimal control algorithm to drive the system from 
the initial state $|\Psi_{ini}\rangle\equiv|G(0)\rangle\otimes|0\rangle$ to the final state 
$|\Psi_{fin}\rangle\equiv|G(0)\rangle\otimes|1\rangle$, using the Hamiltonian (\ref{eq:H3m}) for the time 
evolution where $|G(0)\rangle$ is the ground state of the Hamiltonian (\ref{eq:H3}) at
time $t=0$.
The initial state $|G(0)\rangle\otimes|m=0\rangle$ can be achieved
by suddenly coupling the array to the
electrodes. The normalized gate charges $q_{1}$ and $q_{2}$ will be the control parameters.
In all of the numerical simulations we performed we assume $\varphi=0$
so that supercurrent is zero. Moreover, in the adiabatic case and  $\varphi=0$ the
errors are most severe, so that our settings describe the worst case scenario
to test optimal control theory. The expectation value of the counter operator 
${\hat{m}} \equiv\sum_{m} m~|m\rangle\langle m|$ represents the pumped charge $Q_p$ in unit of $2e$. 
An important measure of the accuracy of the pump is given by the deviation of the pumped charge from the quantized value
\begin{equation}
{\cal E} =|1 - \frac{Q_p}{2e}| \;\;. 
\end{equation}
In addition, since Cooper pair pumping is a coherent process, it is also interesting to test the accuracy 
of the optimization protocol by measuring the departure of the final quantum state from the desired 
one. To this end we also study the fidelity at the end of time evolution 
$$
{\cal F}\equiv|(\langle 1|\otimes\langle G(0)|)|\Psi(T)\rangle|^2
$$ 
which measures the overlap between the final state of the array and the ground state $|G(0)\rangle$.
In all plots we show the infidelity defined as ${\cal I} = 1 -{\cal F}$. 

\section{Optimal Cooper pair pumping}
\label{res}
In this section we present our results for the non-adiabatic optimal pump with $\varphi=0$, i. e. 
when there is no supercurrent.
The goal is to achieve a pumped charge as close as possible to $2e$ (we remind that $\varphi=0$
 is the case where the error is maximal without optimization).

\subsection{Optimal pumps}
\label{optimal}

Motivated by the adiabatic case, as initial guess for control parameters we choose a circular path in 
$(q_{1},q_{2})$ space, around the degeneracy point $q_{1}=q_{2}=\frac{1}{3}$ with radius $q_0$ described by 
sinusoidal gate voltages $q_{1}=\frac{1}{3}+q_{0}\cos(2\pi t/T+\theta_{0})$ and $q_{2}=\frac{1}{3}+q_{0}
\sin(2\pi t/T+\theta_{0})$, where $\theta_0=5\pi/4$.
 The number of excess charges considered for each box is $n = 0, \pm1$, so 
that nine lowest charge states of the whole system are allowed to contribute to pumping.
\begin{figure}
\begin{center}
\tabcolsep=0cm
\begin{tabular}{cc}
\includegraphics[width=0.8\linewidth]{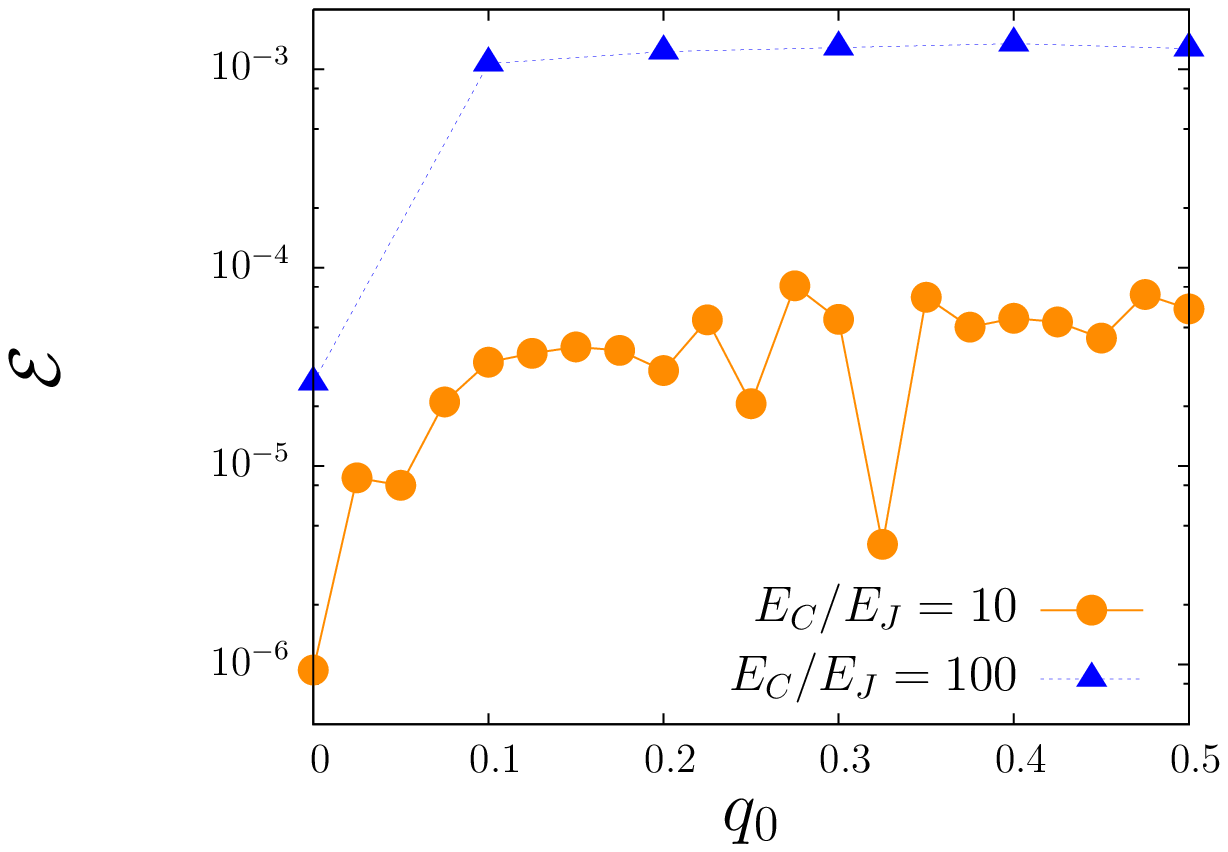}\\
\includegraphics[width=0.8\linewidth]{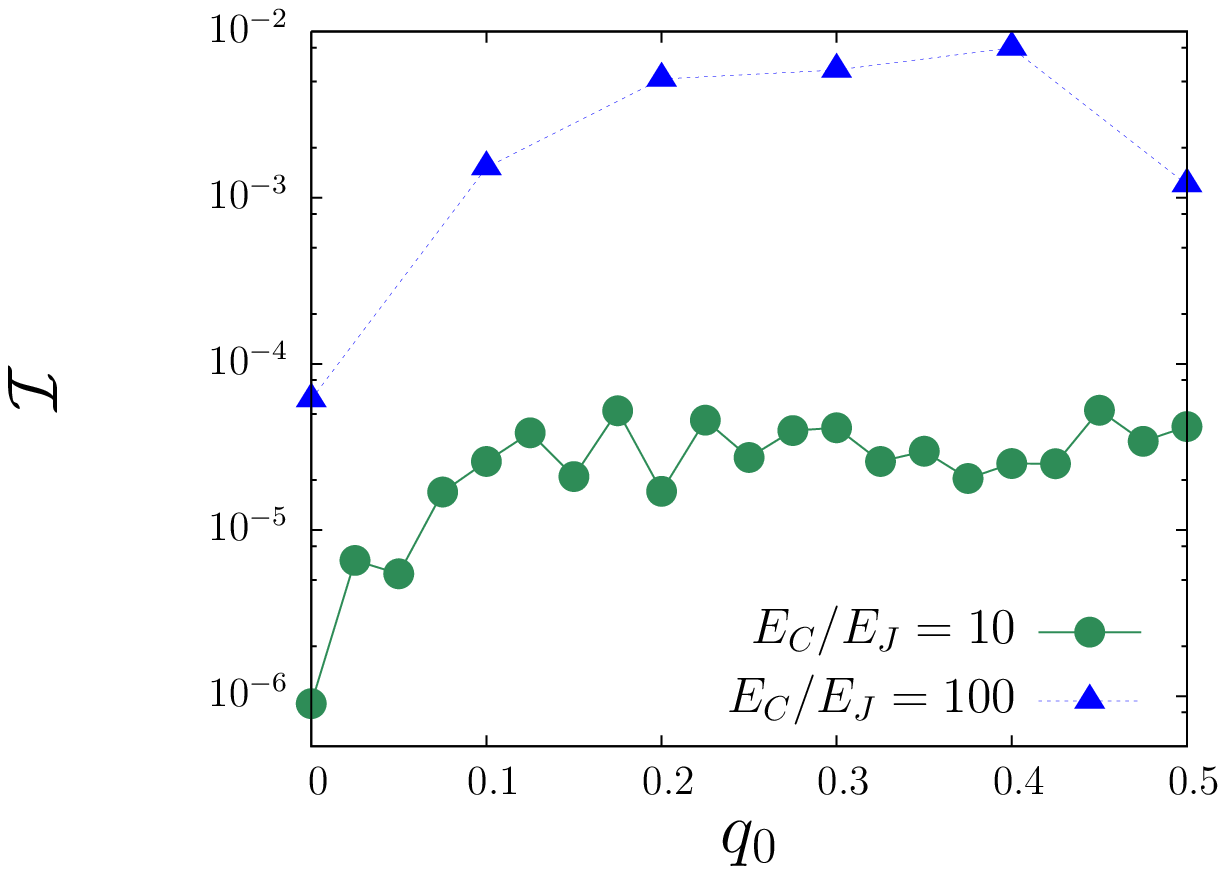}
\end{tabular}
\end{center}
\caption{(Color on line) Top panel: Error in pumped charge ${\cal E}$ and  Bottom panel: Infidelity ${\cal I}$ 
after one cycle as a function of radius of circular path $q_{0}$ in $(q_{1},q_{2})$ space, for $E_C/E_J=10$ 
and $100$. The vertical axis in both panels is in logarithmic scale. Results are obtained after using the 
quantum optimal control theory to modify the circular path (at least 200 iterations). $T=30\pi \hbar/E_{J}$ and 
$\varphi=0$. Numerical error, due to discretization of the time evolution, is at most of the order of $10^{-5}$.
\label{fig:no-noise}}
\end{figure}
Since the initial paths with radii greater than $1/2$ might involve more states, we take $q_0\leq 1/2$.
For each given radius, optimal control algorithm modifies the initial path such that the fidelity 
$\cal F$ after one cycle approaches the value one.
In the following we shall mostly focus on an experimentally relevant case and take $E_{C}/E_{J}=10$. 
In this case in the adiabatic regime the accuracy of the pumped charge is always smaller than fifty 
percent~\cite{pekola99}.
As shown in Fig.~\ref{fig:no-noise}, accurate pumping could be achieved with $T=30\pi \hbar/E_{J}$, which 
is already in the non-adiabatic regime, after at least $200$ iterations.
In the top panel of Fig.~\ref{fig:no-noise} we plot the error in pumped charge ${\cal E}$ as a function 
of radius $q_{0}$, while the bottom panel shows the infidelity ${\cal I}$.
Both error and infidelity are always less than $10^{-4}$.
In Fig.~\ref{fig:no-noise} we show also some results for $E_C/E_J=100$ (triangles),
 with error and infidelity always smaller than $10^{-2}$.
  The difference between the accuracy of these two cases is due to the stronger violation of 
  adiabatic condition, $T\gg \hbar E_C/E_J^2$, in the case $E_C/E_J=100$ compared to $E_C/E_J=10$. 
Notice, moreover, that the optimization procedure allows to control directly the fidelity 
but only indirectly the value of pumped charge so that the behavior of ${\cal I}$ and ${\cal E}$ as 
a function of $q_0$ are not expected to be equal. 
\begin{figure}
\begin{center}
\includegraphics[width=1.0\linewidth]{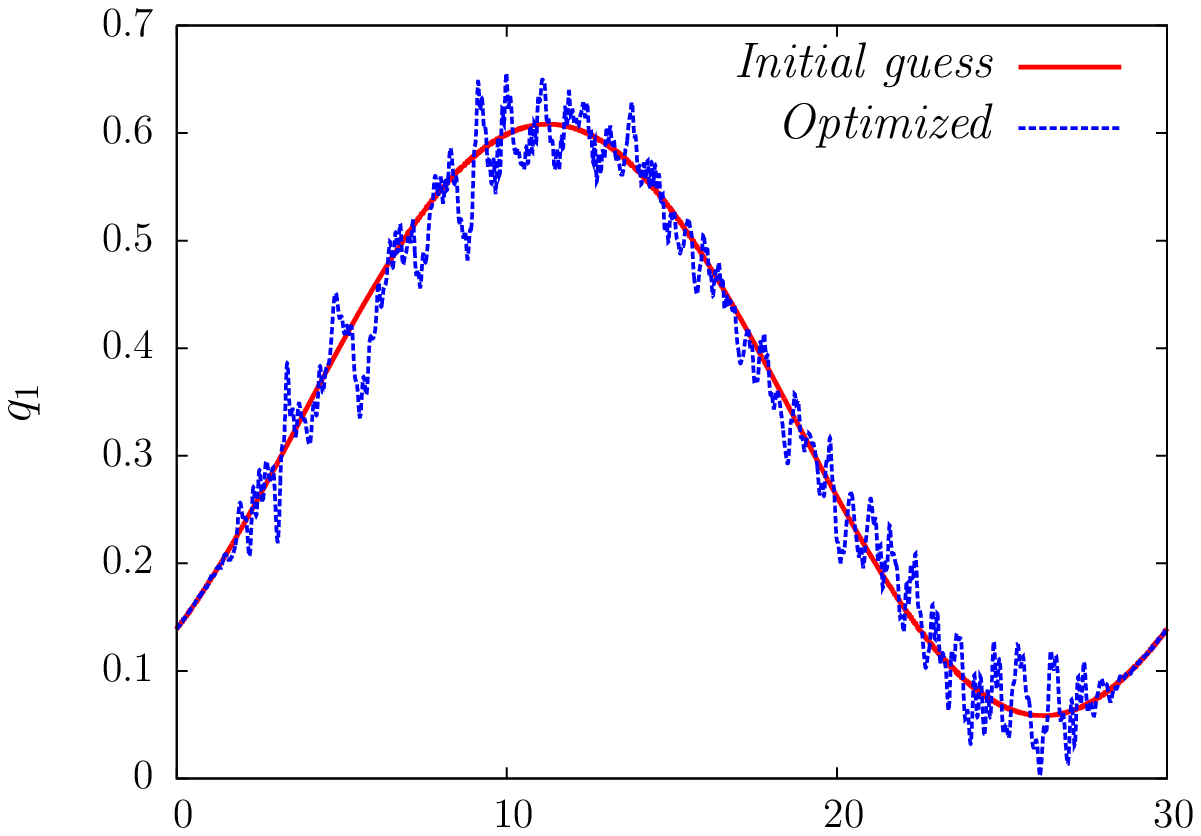}\\
\includegraphics[width=1.0\linewidth]{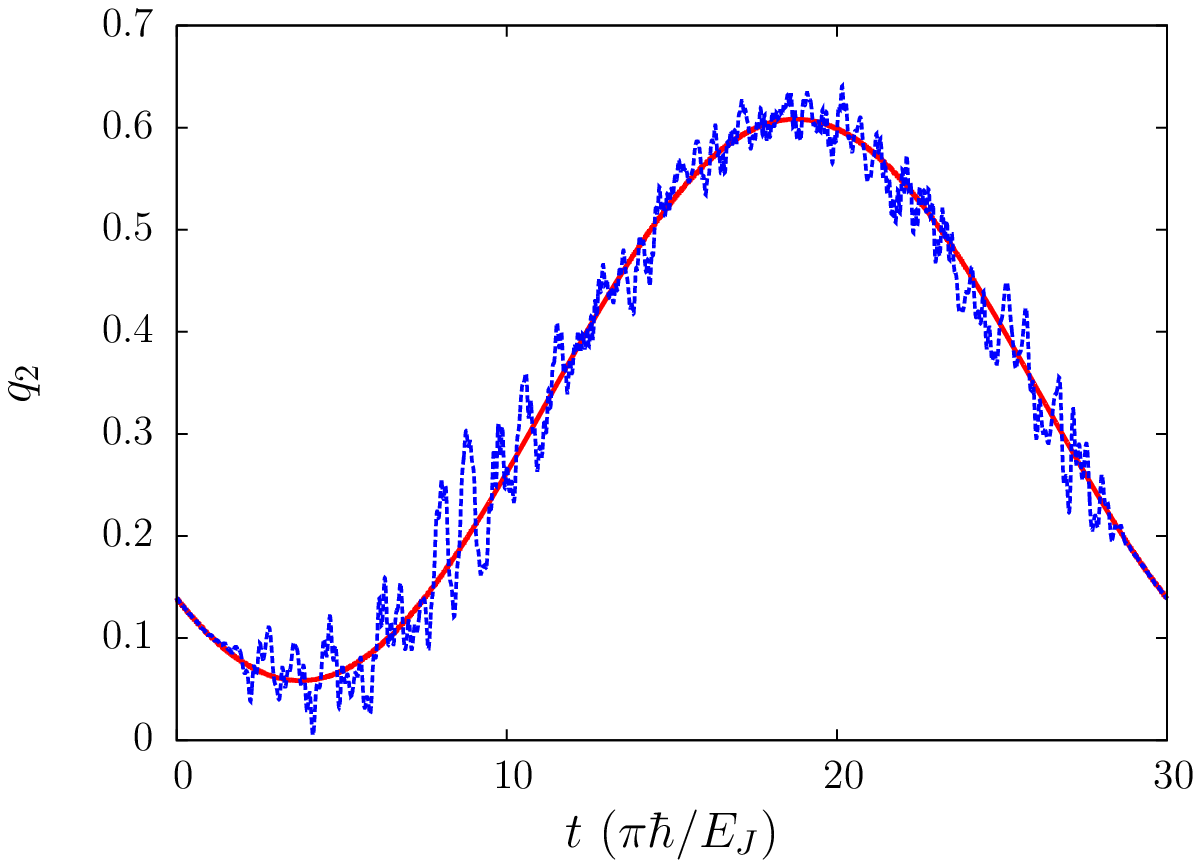}
\caption{(Color on line) An example of gate voltages $q_1$ (top panel) and $q_2$ (bottom panel) as a function 
of time, before and after optimization (200 iterations). The solid red lines correspond to initial circular 
path with radius $q_{0}=0.275$ around the degeneracy point $q_1=q_2=1/3$ in $(q_1,q_2)$ space. Dotted blue 
lines are optimized voltages leading to accurate pumping. $T=30\pi\hbar/E_J$, $E_{C}/E_{J}=10$ and $\varphi=0$. 
\label{fig:pulses}}
\end{center}
\end{figure}
\begin{figure}
\begin{center}
\includegraphics[width=1.0\linewidth]{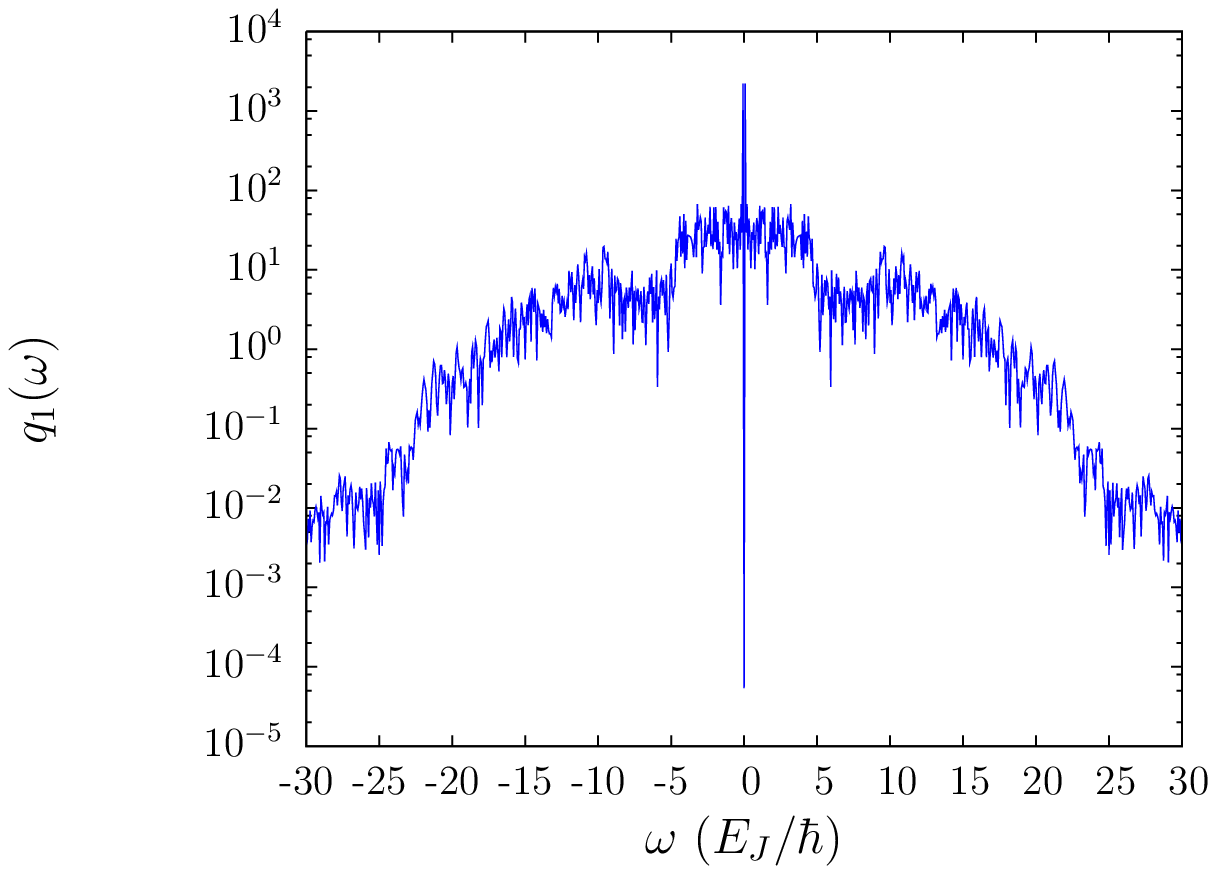}
\includegraphics[width=1.0\linewidth]{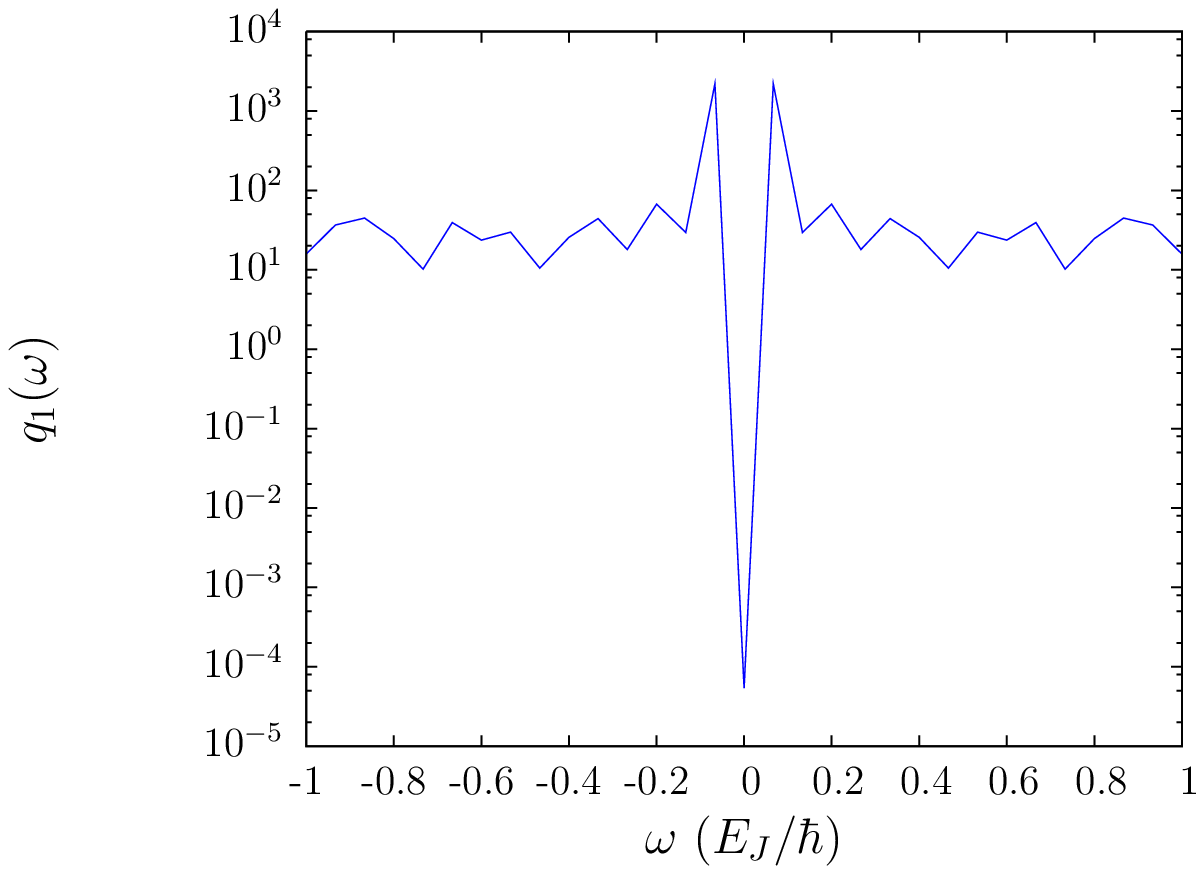}
\caption{(Color on line) An example of Fourier transform of optimized gate voltage $q_1$ 
(blue curve in the top panel of fig.\ref{fig:pulses}). Top panel shows most important harmonics as a function
 of frequency and few first harmonics are shown in bottom panel. $q_{0}=0.275$, $T=30\pi\hbar/E_J$, 
$E_{C}/E_{J}=10$ and $\varphi=0$. 
\label{fig:fourier}}
\end{center}
\end{figure}
\subsection{Imperfections in the pulse shapes}
\label{shapes}
\begin{figure}
\begin{center}
\includegraphics[width=1.0\linewidth]{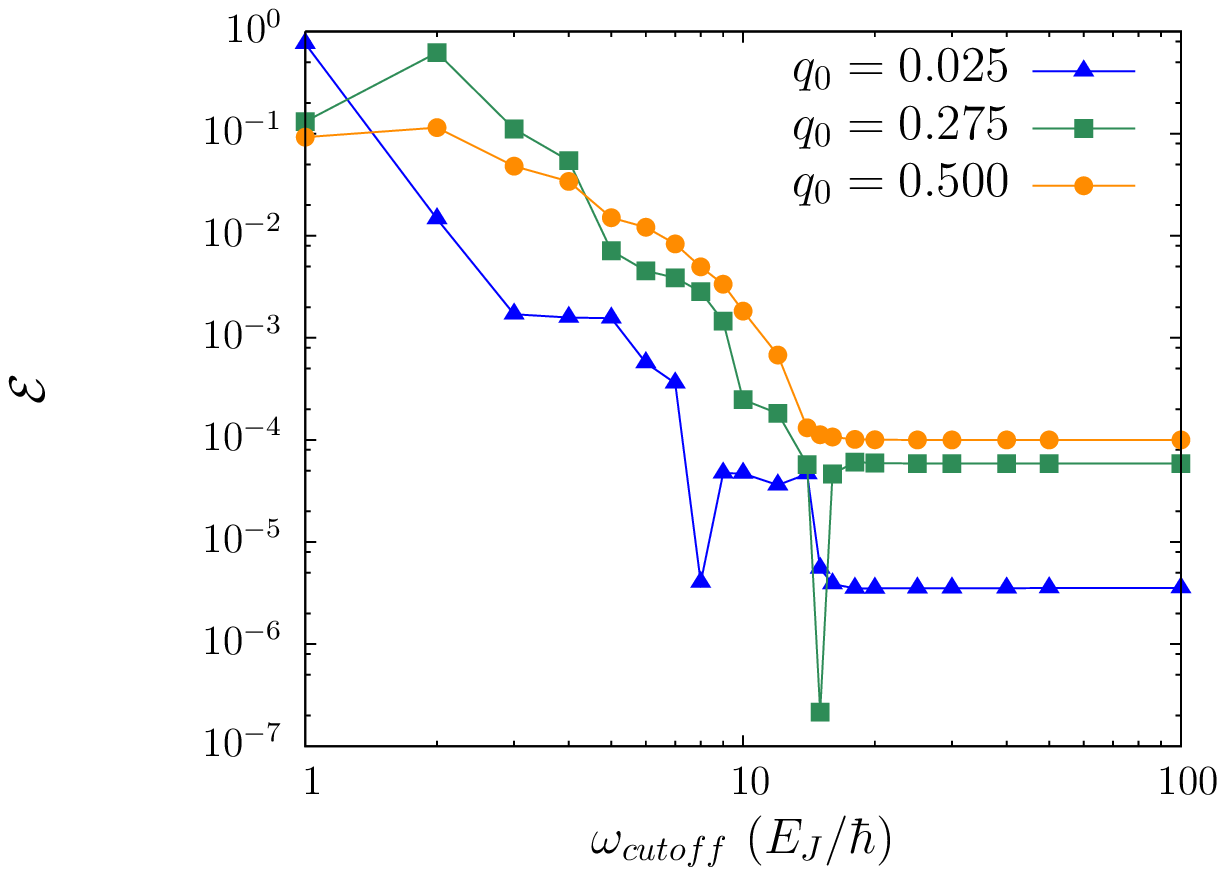}\\
\includegraphics[width=1.0\linewidth]{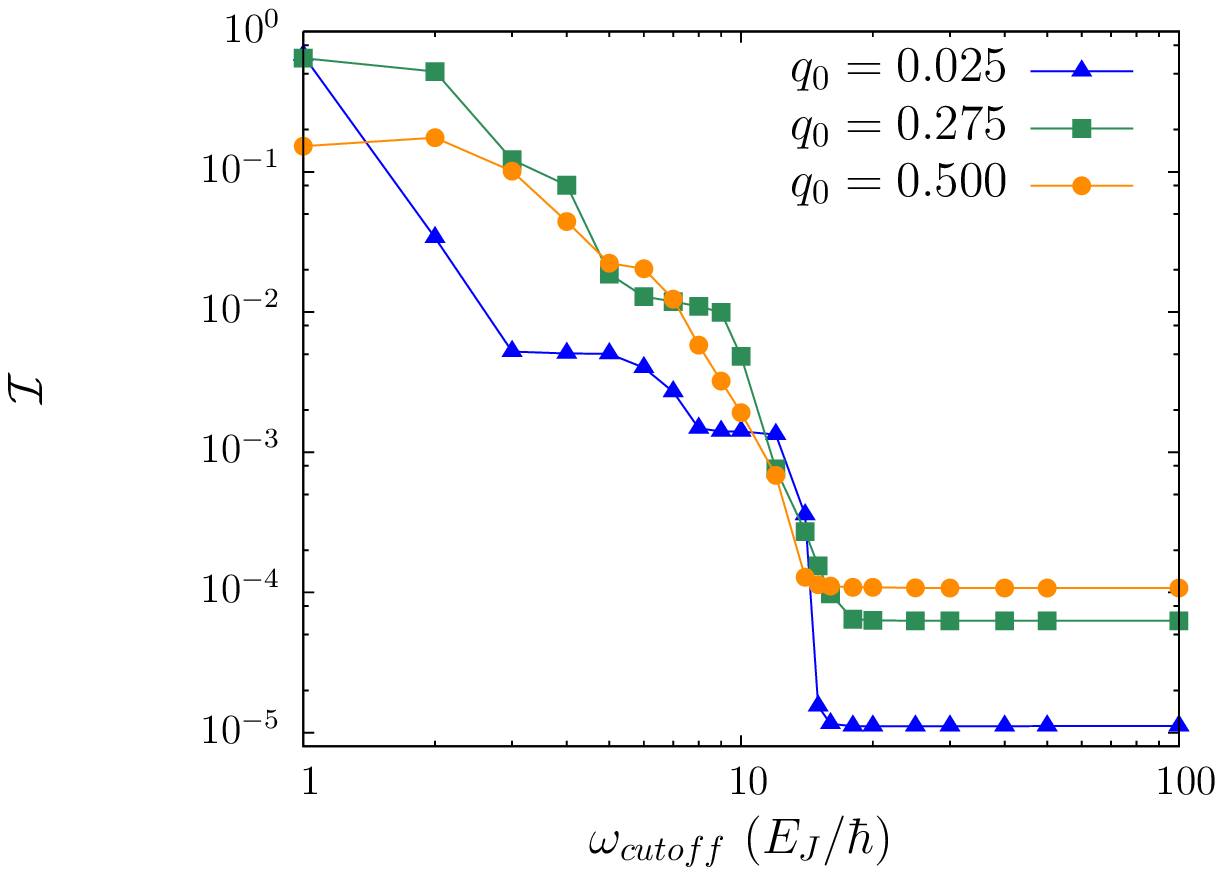}
\caption{(Color on line) The error $\cal E$ in pumping one Cooper pair in each cycle (top panel) and the 
infidelity $\cal I$ of ending up to the ground state of the array at the end of cycle (bottom panel), in 
optimized non-adiabatic case, as a function of the cutoff frequency $\omega_{cutoff}$ of gate voltages. 
Different curves are related to different radii of circular path in $(q_{1},q_{2})$ space as the 
initial guess for control parameters in quantum optimal control theory. $E_C/E_J=10$, $T=30\pi \hbar/E_J$ 
and $\varphi=0$. Numerical error is at most of the order of $10^{-5}$.
\label{fig:cutoff}}
\end{center}
\end{figure}

It is important to understand to which extent the pulses which optimize the pumping can be realized in 
experiments. Fig.~\ref{fig:pulses} shows an example ($q_0=0.275$) of initial (red line) and optimized 
(blue dotted line) gate voltages $q_1$ (top panel) and $q_2$ (bottom panel) after 200 iterations.
In the top panel of Fig.~\ref{fig:fourier} we show the Fourier transform $q_1(\omega)$ of the optimized
 gate voltage $q_1(t)$ plotted in Fig.~\ref{fig:pulses}, while in the bottom panel the same 
curve is plotted on a smaller range of frequencies.
Note that the largest contribution occurs for $\omega=\pm 1/15 E_J/\hbar$, which is the frequency 
of the initial sinusoidal gate voltages.
In a realistic situation, however, it is very difficult to imagine that all the details of the pulse, 
encoded in the high frequency components, can be reproduced faithfully. We accounted for imperfections
in the pulse shape by introducing a bandwidth parametrized by a high frequency cutoff $\omega_{cutoff}$.
In Fig.~\ref{fig:cutoff} we show both the error and the infidelity as a function of the bandwidth 
for three different radii of initial circular path. No changes occur by decreasing $\omega_{cutoff}$ 
from 100 $E_J/\hbar$ until $\omega_{cutoff}\approx 15 E_J/\hbar$ is reached and at this point a 
dramatic increase in both error and infidelity occurs which demonstrates the importance of 
harmonics with frequencies close to this value. Another change in pumped charge and fidelity 
happens when $\omega_{cutoff}\approx5 E_J/\hbar$ is reached. Therefore the most important harmonics 
for optimizing the gate voltages lie below $\omega=20 E_{J}/\hbar$ and there is no need for 
higher frequencies in order to pump accurately.
\subsection{Effect of noise}
\label{noise}
\begin{figure}
\begin{center}
\includegraphics[width=1.0\linewidth]{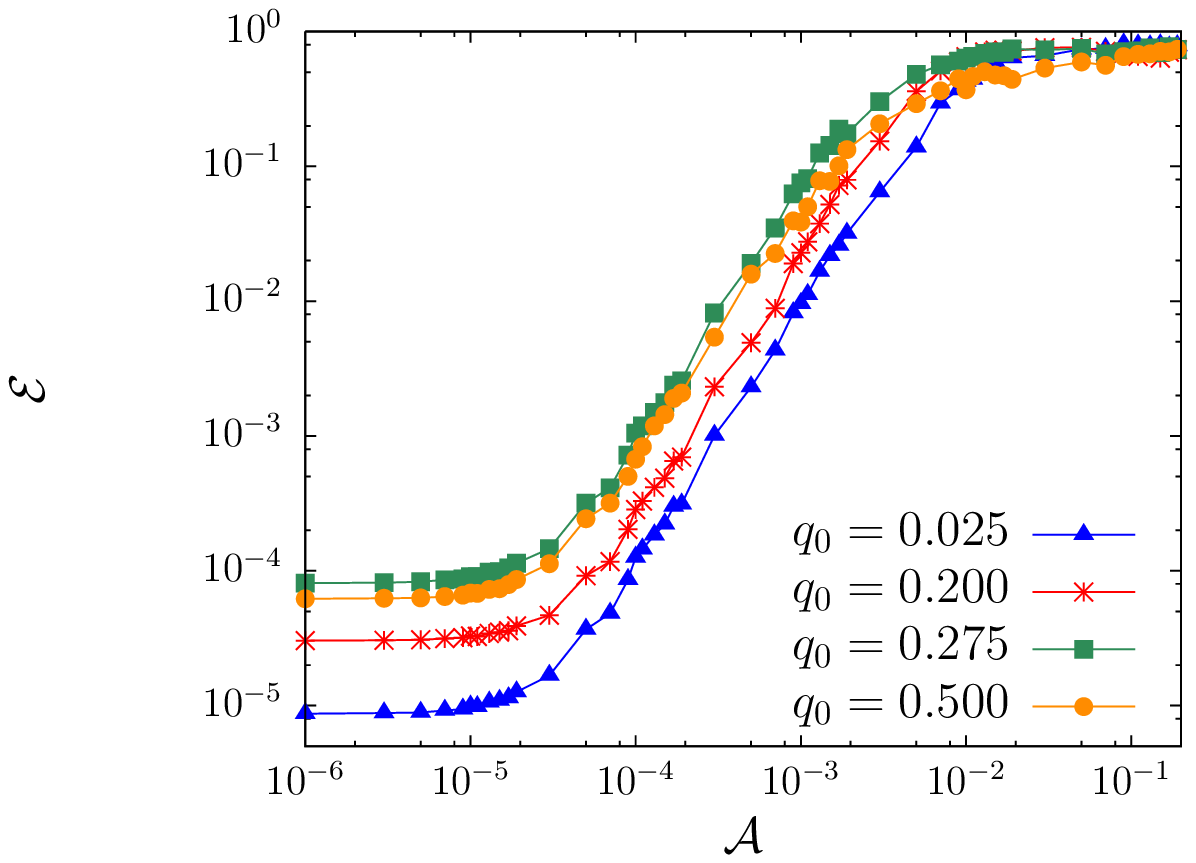}
\includegraphics[width=1.0\linewidth]{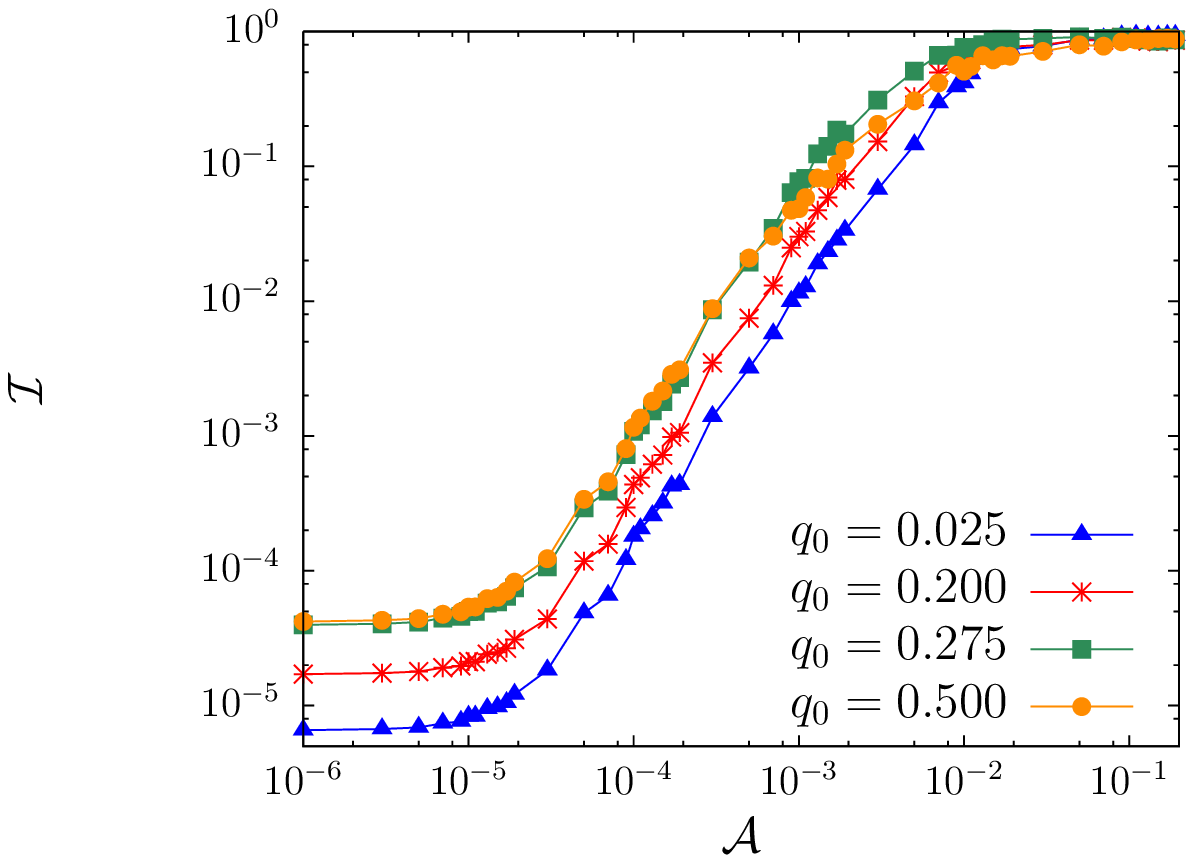}
\caption{(Color on line) The error $\cal E$ in pumping one Cooper pair in each cycle (top panel) and the 
infidelity $\cal I$ of ending up to the ground state of the array at the end of cycle (bottom panel), in 
optimized non-adiabatic case, as a function of the strength of the $1/f$ noise $\cal A$ affecting gate 
voltages. Different curves are related to different radii of circular path in $(q_{1},q_{2})$ space as the 
initial guess for control parameters in quantum optimal control theory. $E_C/E_J=10$, $T=30\pi \hbar/E_J$ 
and $\varphi=0$. Numerical error is at most of the order of $10^{-5}$.
\label{fig:different-noise}}
\end{center}
\end{figure}

We finally discuss the effect of external noise. Since Cooper pair pumping is a coherent process, the presence
of an external environment may be disruptive. In Josephson nanocircuits in the charge regime the dominant 
mechanism of decoherence is $1/f$ noise (see e.g. Ref.~\cite{ithier05}). 
It is important to know whether the pumped charge is stable while optimized gate voltages are affected by 
noise. Although its understanding is far from complete, $1/f$ noise is believed to originate from two-level 
fluctuators present in the substrate and/or in the insulation barrier. Several theoretical works have recently 
studied the effect of $1/f$ noise~\cite{backgroundcharges}. 
Following current approaches, we model the environment as a superposition of bistable 
classical fluctuators resulting in an additional random contribution $\delta q_{k}(t)$ to the gate 
charges after optimization.
\begin{figure}
\begin{center}
\tabcolsep=0cm
\begin{tabular}{cc}
\includegraphics[width=0.8\linewidth]{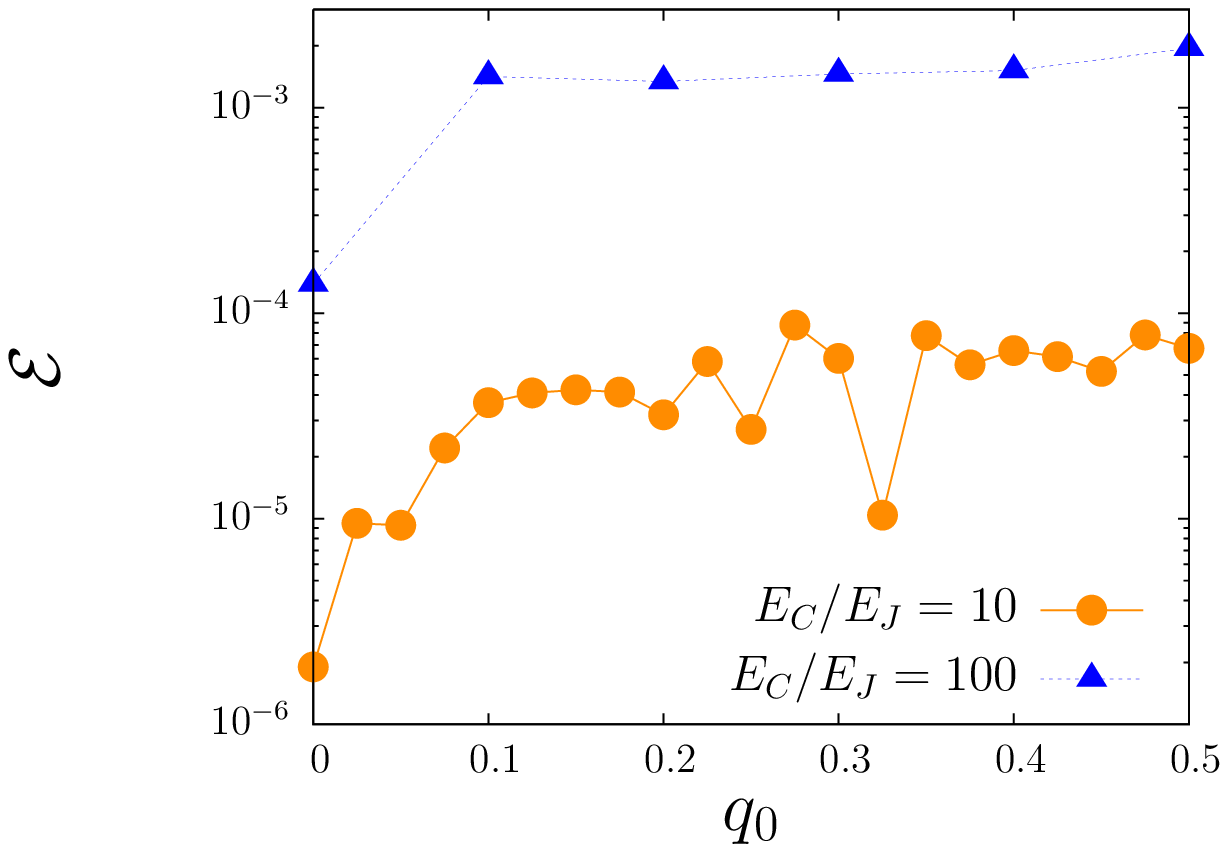}\\
\includegraphics[width=0.8\linewidth]{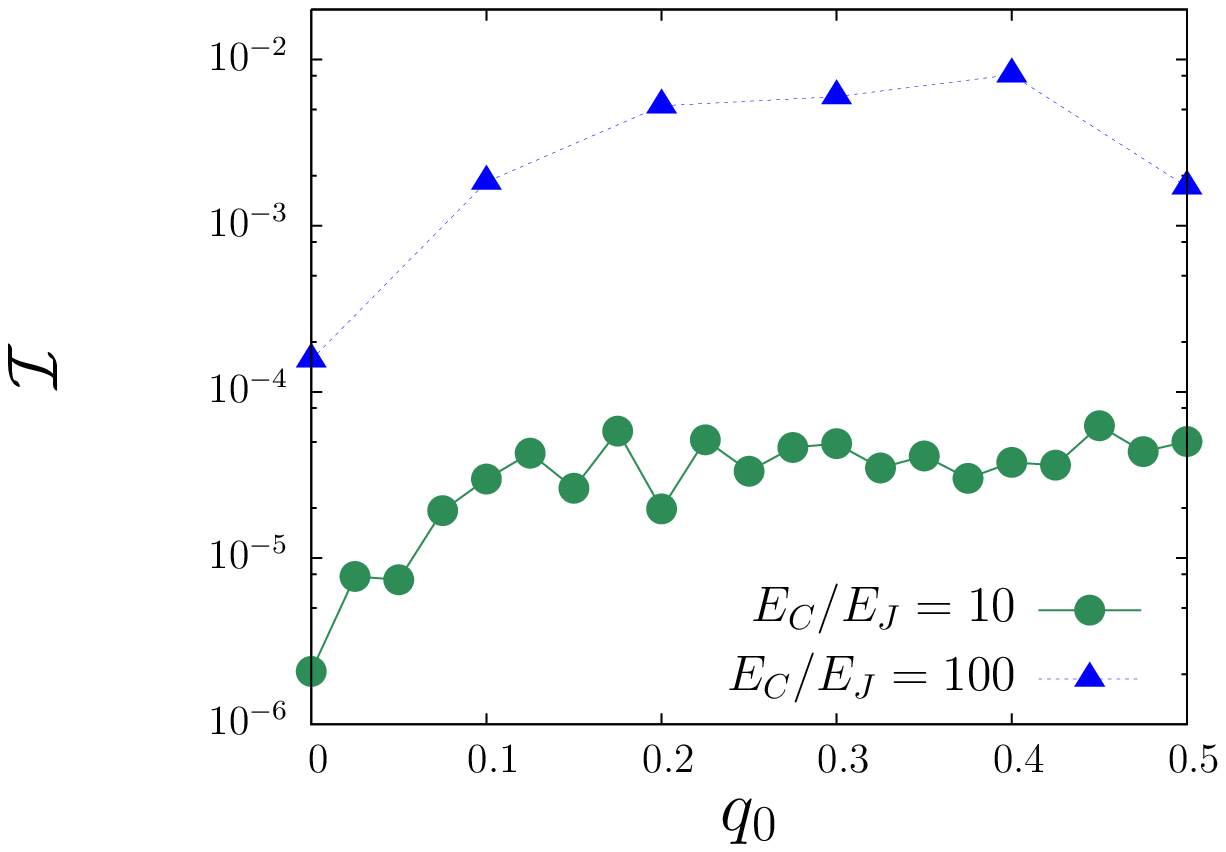}
\end{tabular}
\end{center}
\caption{(Color on line) Error in pumped charge ${\cal E}$ (top panel) and infidelity ${\cal I}$ (bottom panel) 
after one cycle as a function of radius of circular path $q_{0}$ in $(q_{1},q_{2})$ space, for $E_C/E_J=10$ and 
$100$, after applying the noise with power spectrum ${\cal S}_{q_k}(\omega)={\cal A}/\omega$ and strength 
${\cal A}=10^{-5}$ on optimized gate voltages. $T=30\pi\hbar/E_{J}$ and $\varphi=0$. The vertical axis in 
both panels is in logarithmic scale. Numerical error is at most of the order of $10^{-5}$.
\label{fig:with-noise}}
\end{figure}
 A distribution of switching rates $\gamma$ behaving as $P(\gamma)\propto 1/\gamma$ 
in a range $[\gamma_{min}:\gamma_{max}]$ results in a noise power spectrum ${\cal S}_{q_k}(\omega)=
\langle\delta q_{k}(t)\delta q_{k}(0)\rangle_{\omega}\approx\omega^{-1}$.
We chose the switching rates such that the $1/f$ part of the spectrum is centered around the typical 
frequency of the pump. Typically the $1/f$ region extends over two order of magnitudes in $\omega$.
We assume that fifty independent fluctuators are coupled weakly to the system and that the charge 
noise on the two separate gates are uncorrelated.
Moreover we averaged the results over fifty different configurations of noise.
In Fig.~\ref{fig:with-noise} we plot the error $\cal E$ (top panel) and infidelity $\cal I$ (bottom panel), 
as a function of $q_0$, for a strength of noise ${\cal A}=10^{-5}$ (${\cal S}(\omega)={\cal A}/\omega$), 
which is a typical experimental value.
Remarkably, the pumped charge is virtually unaffected by noise: for $E_C/E_J=10$ the error in the pumped 
charge and infidelity are still less than $10^{-4}$, which clearly shows 
the stability against noise. The case $E_C/E_J=100$ is also stable.
To complete the analysis, we plot in Fig.~\ref{fig:different-noise} the error in pumped charge (top panel) 
as well as the infidelity (bottom panel) as a function of $\cal A$ for a few different values of $q_0$ 
(note that both vertical and horizontal axes are in logarithmic scale).
For all values of radius $q_0$ the accuracy of pumping and fidelity are larger than $90 \%$ even under 
the effect of noise with significant strength ${\cal A}=10^{-3}$, which is a good improvement considering 
that the accuracy is less than fifty percent in adiabatic regime without noise.
More importantly, both ${\cal E}$ and ${\cal I}$ are not increasing up to about ${\cal A}=10^{-5}$.
\begin{figure}[h]
\begin{center}
\includegraphics[width=0.8\linewidth]{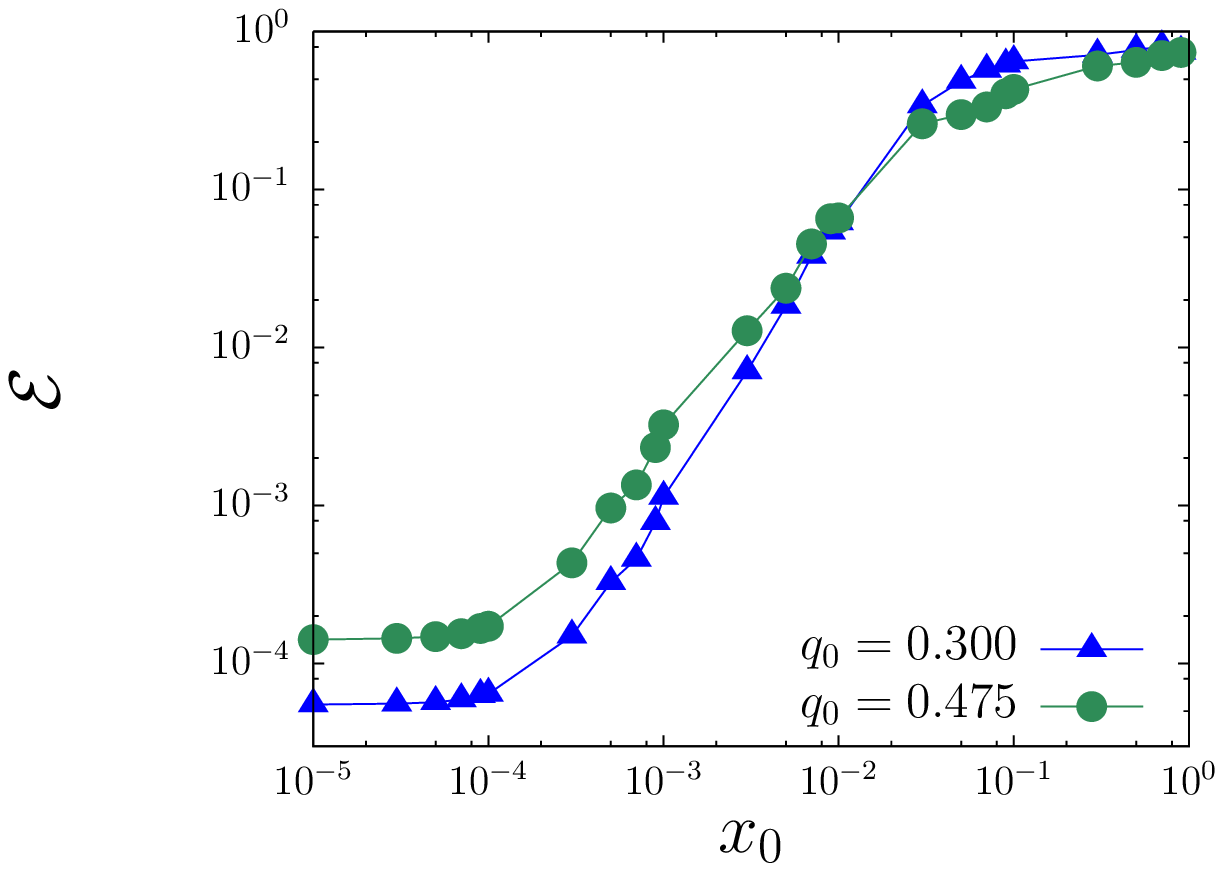}
\includegraphics[width=0.8\linewidth]{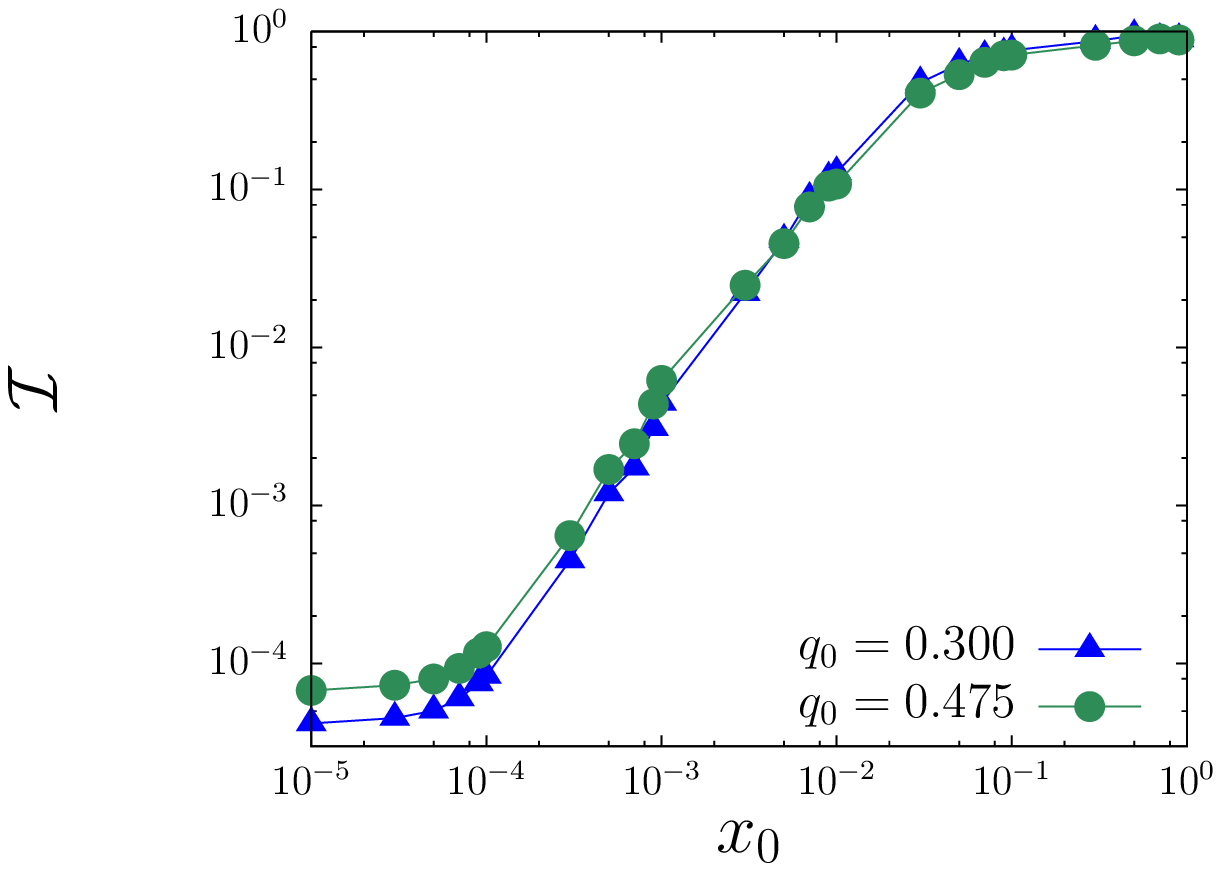}
\caption{(Color on line) The error $\cal E$ (top panel) and the infidelity $\cal I$ (bottom panel), 
as a function of the random fraction $x_0$ added to the nominal $E_{C}$ and $E_{J}$.
 Optimized pulses obtained for uniform array are used to calculate pumped charge and fidelity.
 $E_C/E_J=10$, $T=30\pi \hbar/E_J$ 
and $\varphi=0$. Numerical error is at most of the order of $10^{-5}$.
\label{fig:random-noise}}
\end{center}
\end{figure}
\begin{figure}[h]
\begin{center}
\includegraphics[width=0.8\linewidth]{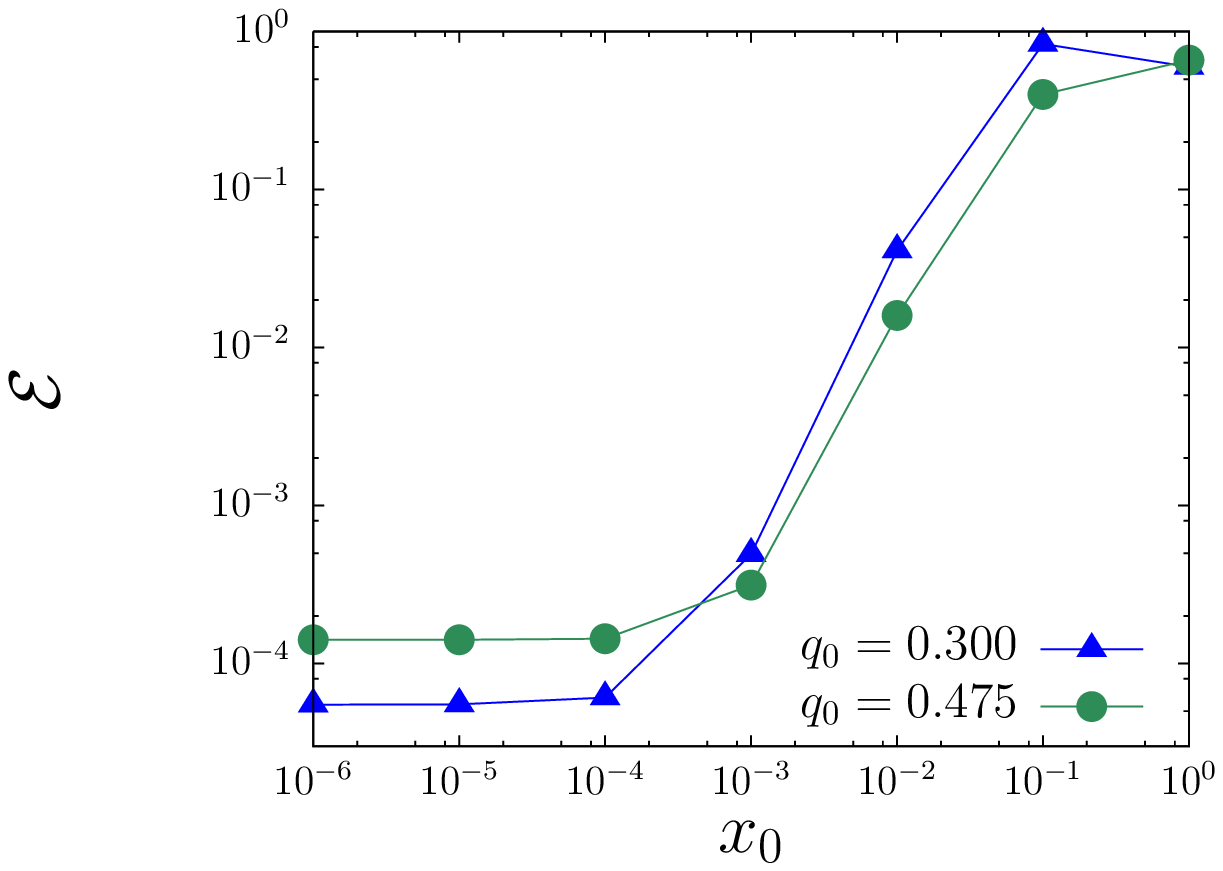}
\includegraphics[width=0.8\linewidth]{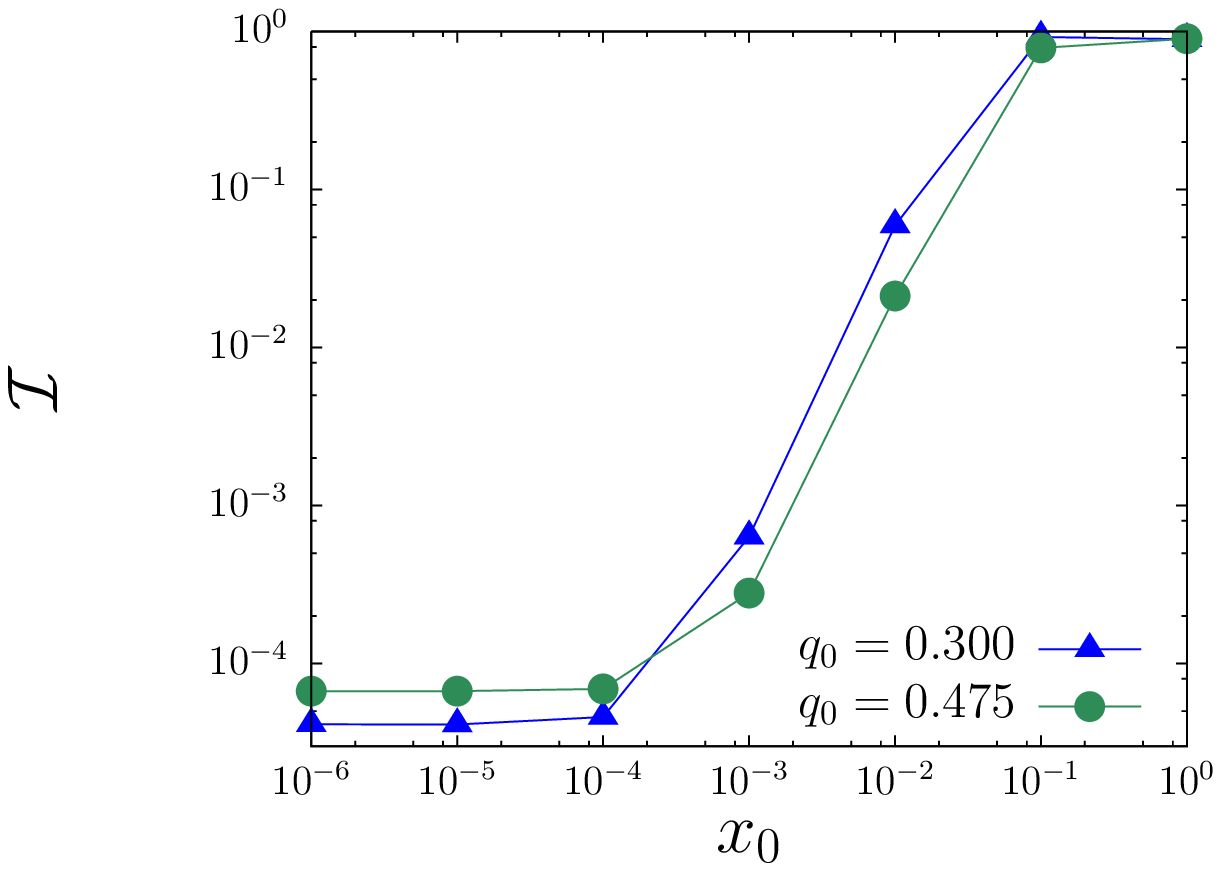}
\caption{(Color on line) The error $\cal E$ (top panel) and the infidelity $\cal I$ (bottom panel), 
as a function of the weight $x_0$ of the random Hamiltonian added to the uniform Hamiltonian.
Optimized pulses obtained from non-perturbed Hamiltonian are used to calculate pumped charge and fidelity.
 $E_C/E_J=10$, $T=30\pi \hbar/E_J$ and $\varphi=0$. Numerical error is at most of the order of $10^{-5}$.
\label{fig:perturbation}}
\end{center}
\end{figure}
\subsection{Imperfections in the pump}
\label{parameters}
Up to now we have assumed the ideal situation in which the parameters of the pump are known. In this section we address the effect of imperfections arising when such parameters are known only up to a given uncertainty from a uniform array.
More precisely, we estimate how error in pumped charge and infidelity increase, as a function of the extent of the uncertainty, by evolving the system using the optimized pulse shapes calculated for the uniform array. To assess this issue we have implemented such an uncertainty on the parameters with two methods. In the first one we add to the nominal $E_C$ and $E_J$, of a uniform array, a random fraction $x$ taken in the range $[-x_0,x_0]$. In Fig. \ref{fig:random-noise} we plot error and infidelity, respectively, averaged over 50 configurations as a function of $x_0$ for $E_C=10E_J$, for $q_0 = 0.300$ and $q_0 = 0.475$. In the second method we perturb the uniform Hamiltonian by adding to it a weighted random matrix. In Fig. \ref{fig:perturbation} we plot error and infidelity, respectively, as a function of the weight $x_0$ for $E_C=10E_J$ and for two values of $q_0$. Both methods show that no increase of infidelity and epsilon is found up to a fraction (or a weight) of the order of $10^{-4}$. 
Finally, we emphasize that we have always assumed a uniform array simply for the sake of definiteness. Indeed, in the case of non-uniformity of the parameters, as long as they are known, the pulse optimization can always be performed. Of course, the shape of the final optimized pulses would be different, with respect to the uniform situation, depending on the extent of the non-uniformity.
\section{Conclusion}
In this paper we presented a proposal for realizing fast and accurate Cooper pair pumps. The idea was to 
use a conventional gate-based Cooper pair pump with optimized pulse shapes. This pump operates
in the non-adiabatic regime and, as we showed, can achieve a high accuracy. Numerical results after using
 the optimal control theory were presented demonstrating that pumped charge is accurate and stable against 
the noise applied on gate voltages and uncertainty in charging and Josephson energy.
 Moreover all important harmonics contributing to optimized gate voltages have 
frequencies less than few $E_J/\hbar$. The simple design of the pump, the short time scales of operation  
and the stability against noise are advantages which make it worthy to think of the quantum optimal control 
theory as a powerful tool to achieve more realistic and still accurate pumping. As far as experiments are concerned,
 the limits on the bandwidth of the pulse generator are still demanding. 

In order to apply optimal control theory we enlarged the Hilbert space to take into account a passive 
detector which acts as a counter. We assumed that initially the counter was set to zero implying that 
we supposed to disconnect the superconducting network from the electrodes. Moreover,
 we concentrated only on the case $\varphi=0$.  It would be important to find other 
methods for optimization which do not need to introduce counter so to explore also phase 
dependent errors. 
\acknowledgments
We would like to acknowledge very fruitful discussions with T. Calarco
and V. Brosco. We acknowledge support 
by EC-FET/QIPC (EUROSQIP).


\begin{references}
\bibitem{footnote}
	In superconducting systems, in addition to pumped charge, there is another contribution
	to current due to the Josephson effect if the superconducting phase difference between the two superconducting 
	electrodes is not zero.
\end{references}
\end {document}